\documentclass[letterpaper, 10 pt, conference]{ieeeconf}  

\IEEEoverridecommandlockouts                              

\usepackage{cite}
\usepackage{amsmath,amssymb,amsfonts}
\usepackage{algorithmic}
\usepackage{booktabs}
\usepackage{graphicx}
\usepackage{textcomp}
\usepackage{xcolor}
\usepackage{etoolbox}
\usepackage{graphicx}
\usepackage[skip=0.333\baselineskip]{caption}
\usepackage{subcaption}
\usepackage{multirow}
\usepackage{multicol}
\usepackage{tabularx}
\usepackage{hyperref}
\usepackage{balance}
\usepackage{graphicx}
\usepackage{subcaption}
\usepackage{adjustbox}

\title{\LARGE \bf
Listen, Look, Drive: Coupling Audio Instructions for User-aware VLA-based Autonomous Driving
}

\author{
   Ziang Guo$^{\dagger}$, Feng Yang$^{\circ}$, Xuefeng Zhang$^{\dagger}$, Jiaqi Guo$^{\dagger}$, Kun Zhao$^{\dagger}$, \\ Yixiao Zhou$^{\ddag}$, Peng Lu$^{\circ}$, Sifa Zheng$^{\S}$, Zufeng Zhang$^{\dagger \star}$
    \thanks{$^{\dagger}$Ziang Guo, Xuefeng Zhang, Jiaqi Guo, Kun Zhao, and Zufeng Zhang are with SuZhou Automotive Research Institute of Tsinghua University, China. The work is done during Ziang Guo's internship in SuZhou Automotive Research Institute of Tsinghua University.
    \tt slxx5237@gmail.com, osu\_zxf@126.com, guojiaqi9316@gmail.com, \{zhaokun, zhangzufeng\}@tsari.tsinghua.edu.cn}
    \thanks{$^{\ddag}$Yixiao Zhou is with Department of Electrical and Electronic Engineering, The University of Hong Kong, China. \tt u3649491@connect.hku.hk}
    \thanks{$^{\circ}$Feng Yang and Peng Lu are with Hyundai Motor Advanced Tech. R\&D Center, China (HMATC).  
    \tt Feng.Yang@hyundai.com, lupeng@hyundai-hmatc.com}
    \thanks{$^{\S}$Sifa Zheng is with School of Vehicle and Mobility, Tsinghua University. \tt zsf@tsinghua.edu.cn}
    \thanks{$^{\star}$Corresponding Author: Zufeng Zhang}
}
\newif\ifanonymous

\begin{document}

\maketitle
\thispagestyle{empty}
\pagestyle{empty}

\begin{abstract}
Vision Language Action (VLA) models promise an open-vocabulary interface that can translate perceptual ambiguity into semantically grounded driving decisions, yet they typically treat language as a static prior fixed at inference time. As a result, the model must infer continuously shifting objectives from pixels alone, yielding delayed or overly conservative maneuvers. We argue that effective VLAs for autonomous driving need an online channel in which users can influence driving with specific intentions. To this end, we present EchoVLA, a user-aware VLA that couples camera streams with in situ audio instructions. We augment the nuScenes dataset with temporally aligned, intent-specific speech commands generated by converting ego-motion descriptions into synthetic audios. Further, we compose emotional speech-trajectory pairs into a multimodal Chain-of-Thought (CoT) for fine-tuning a Multimodal Large Model (MLM) based on Qwen2.5-Omni. Specifically, we synthesize the audio-augmented dataset with different emotion states paired with corresponding driving behaviors, leveraging the emotional cues embedded in tone, pitch, and speech tempo to reflect varying user states, such as urgent or hesitant intentions, thus enabling our EchoVLA to interpret not only the semantic content but also the emotional context of audio commands for more nuanced and emotionally adaptive driving behavior. In open-loop benchmarks, our approach reduces the average L2 error by $59.4\%$ and the collision rate by $74.4\%$ compared to the baseline of vision-only perception. More experiments on nuScenes dataset validate that EchoVLA not only steers the trajectory through audio instructions, but also modulates driving behavior in response to the emotions detected in the user's speech.
\end{abstract}

\section{Introduction}
\begin{figure}
    \centering
    
    \includegraphics[width=0.98\linewidth]{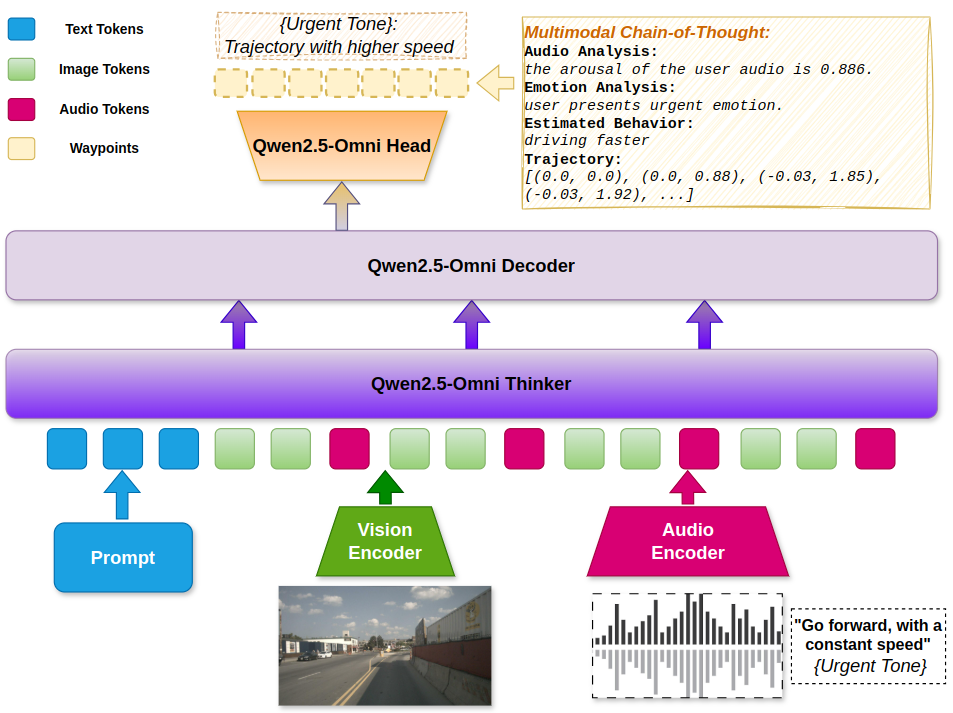}
    \caption{Framework overview of EchoVLA. In EchoVLA, we design a multimodal CoT reasoning to equip VLA with audio-instructed action output, user emotion detection, and emotion-guided trajectory modulation.}
    \label{fig:overview}
\vspace{-5mm}
\end{figure}

Along with the vigorous development of Large Language Models (LLMs), Vision Language Models (VLMs) have shown impressive potential in multimodal understanding \cite{bai2025qwen2, yao2024minicpm, li2024llava}. By jointly encoding camera streams and natural language, modern VLMs translate perceptual ambiguity into semantically grounded representations, offering autonomous driving pipelines an open-vocabulary interface that no longer relies on hand-mapped object catalogs \cite{tian2024drivevlm, jiang2025alphadrive}. However, current VLMs still treat language as a static prior: the prompt is fixed at inference time and cannot adapt when the user suddenly demands. Because the visual scene is intrinsically continuous and the goal hypotheses evolve every frame, the model must infer revised objectives from ambiguous pixels alone, often yielding delayed or overly conservative actions \cite{xie2025ready}. This observation motivates Vision Language Action models (VLAs) to fuse multimodal perception, reasoning, and control into one differentiable sequence-to-sequence problem.

VLAs have been proven to be effective in robotics, where they guide robots through structured quasi-static tasks with well-defined goals \cite{din2025vla-mani, shao2025vla-robo}. Autonomous driving, on the contrary, presents a continuous stream of abstract objectives that vary in time and must be inferred from highly diverse traffic scenes \cite{surmann2025multi}. Most existing autonomous driving VLAs show promising open-loop accuracy, but remain confined to offline captions and lack an online language channel that users can influence in real time \cite{jiang2025surveyVLA}.

To provide VLAs with an explicit, human-interpretable context, we propose adding audio instructions that provide complementary temporal and semantic information. We extend the nuScenes dataset by augmenting each frame with context-relevant audio commands, forming a new image-audio instruction dataset \cite{caesar2020nuscenes}. Then we extend the dataset with emotional user intention and the corresponding trajectories. Using the extended dataset, we perform supervised fine-tuning to a Multimodal Large Model (MLM) in a Chain-of-Thought (CoT) reasoning approach. The fine-tuned MLM is able to receive the vision input from the sensors and the user's audio input to infer user emotion and user-aware trajectory. To validate our approach, we compared the VLA with only visual perception input and our EchoVLA. In nuScense open-loop experiments, EchoVLA achieved a lower L2 error and a lower collision rate. In addition, we also verified the emotional recognition and driving behavior guidance of EchoVLA in the nuScenes dataset.

\section{Related Work}
\begin{figure}
    \centering
    \includegraphics[width=0.85\linewidth]{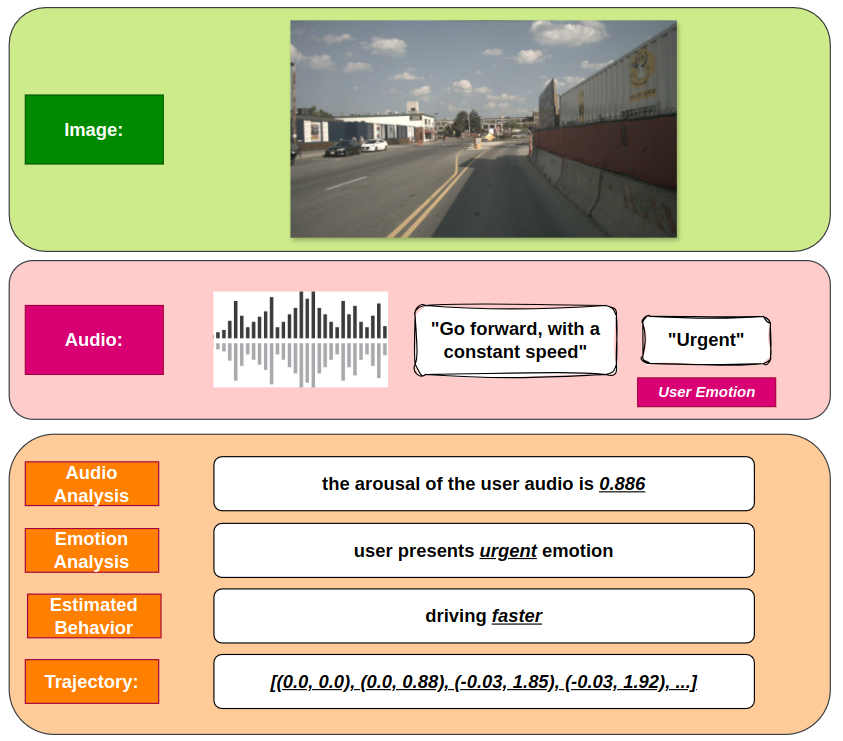}

    \caption{Dataset examples on urgent emotion.}
    \label{fig:dataset}
    \vspace{-6mm}
\end{figure}

\subsubsection{Audio-visual Navigation in Robotics}

On an indoor scale, audio navigation shows impressive progress. DialNav is a collaborative embodied dialog navigation framework in which a navigator agent and a remote guide communicate through multi-turn dialog to reach a target location, underscoring the critical role of conversation for mission success \cite{han2025dialnav}. Huang et al. introduce a novel 3D spatial map representation that integrates audio, visual, and language cues to enable robots to navigate using multimodal queries, leveraging pre-trained multimodal foundation models to store cross-modal information in a centralized 3D voxel grid, allowing for zero-shot spatial goal navigation and improved disambiguation of goal locations. They then introduce multimodal spatial language maps (MSLMs) as a 3D spatial representation that fuses multimodal features with a 3D reconstruction of the environment. The authors propose two instances: Visual-Language Maps (VLMaps) and their extension, Audio-Visual-Language Maps (AVLMaps). These maps enable robots to understand and navigate using natural language commands, images, or audio snippets, even in ambiguous scenarios, showing improved performance in zero-shot spatial and multimodal goal navigation \cite{huang2023avlmaps}.

\begin{figure*}
    \centering
    \includegraphics[width=0.88\linewidth]{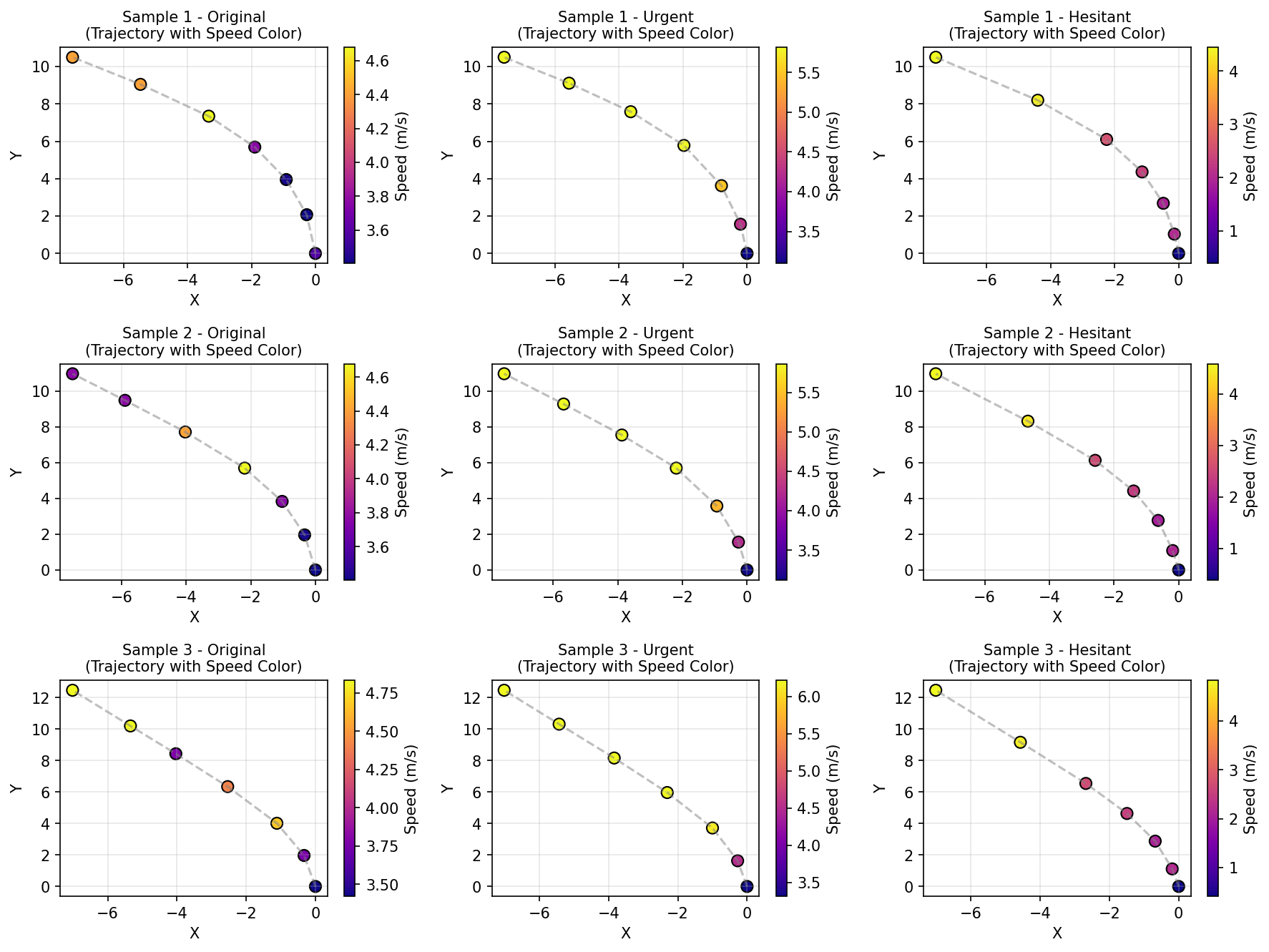}
    \caption{The example of trajectory modulation based on the emotions. For the recognition of user's urgent emotion, the trajectory is modulated with faster speed profile. While under the hesitant emotion, the trajectory is modulated with slower speed profile.}
    \label{fig:emo_traj}
    \vspace{-5mm}
\end{figure*}

\subsubsection{VLM-based Perception and Scene Understanding}

Many powerful vision techniques, such as video encoders and visual prompting, etc., are combined with VLMs to boost VLMs' ability of visual perception and understanding. OmniDrive proposes a causal 3D query formulation that allows an LLM to reason over multiview video without dense BEV rasterisation, improving long-tail detection on nuScenes \cite{wang2024omnidrive}. DriveVLM couples chain-of-thought prompting with hierarchical scene graphs to expose occluded actors and fine-grained relations \cite{tian2024drivevlm}. DriveSOTIF injects VLM-generated semantic masks into an uncertainty estimator, cutting false positive Safety of the Intended Functionality (SOTIF) alerts \cite{huang2025drivesotif}. K Renz et al. introduce an approach that allows the LLM-based driver to interpret driving scenarios by converting object-level vector representations of the environment into human-readable language captions, allowing it to reason about observations and predict actions \cite{renz2024carllava}. Y Hu et al. introduce LLM-RCO, a framework that uses multimodal large language models to integrate human-like common sense into autonomous driving systems, specifically to address partial perception deficits. It aims to improve driving safety and resilience by inferring potential hazards and planning proactive, context-aware control actions, moving beyond traditional immediate stops or minimal risk maneuvers \cite{hu2025combating}. S4-Driver leverages MLMs with a novel sparse volume representation. This approach improves 3D spatiotemporal reasoning by aggregating multiview and multiframe visual input, enabling the model to learn complex driving behaviors without human annotations \cite{xie2025s4}. Integration of powerful vision processing into VLMs in a unified manner boosts the ability to perceive and understand the scene, but leads to significant computation latency. 

\subsubsection{VLM-based Planning and Control}

VLM-MPC is a two-layer architecture in which the upper-layer VLM, composed of an environment encoder, scene encoder, reference memory, and prompt generator, produces key driving parameters such as desired speed and headway based on visual and textual inputs. The lower layer MPC then uses these parameters to control the vehicle in real time \cite{long2024vlm-mpc}. VLMPlanner proposes a hybrid planning framework for autonomous driving that combines a real-time planner with a VLM. This system leverages multiview image data for detailed environmental understanding and introduces a Context-Adaptive Inference Gate (CAI-Gate) to dynamically adjust the VLM inference frequency based on the complexity of the scene, optimizing both performance and computational efficiency \cite{tang2025vlmplanner}. CoT-Drive uses LLMs and a chain-of-thought (CoT) prompting method to improve motion forecasting tasks and employs a teacher-student knowledge distillation strategy to transfer the advanced scene understanding capabilities of powerful LLMs to lightweight language models \cite{liao2025cot}. SimLingo creates a VLA for autonomous driving that unifies language-action alignment by generating various pairs of instruction-action and explicitly training the model to predict actions based on language commands \cite{renz2025simlingo}. AlphaDrive utilizes a GRPO-based RL strategy with four tailored rewards and a two-stage training paradigm that combines supervised fine-tuning with reinforcement learning to improve planning performance and training efficiency, especially in complex long-tail scenarios \cite{jiang2025alphadrive}. RAG-Driver enhances VLMs for autonomous driving by integrating retrieval-augmented generation (RAG) and forming a video memory database to enable action explanation, action justification, and prediction of the next control signal \cite{yuan2024rag}. To adjust the knowledge embedded in VLMs for autonomous driving, extended knowledge base and language-action alignment are common techniques. However, the design of the adjustments remains highly hand-crafted.

\subsubsection{VLM-based Explainability and Interaction}

Careful and task-oriented dataset construction plays a critical role in improving the effectiveness of VLMs. By aligning data collection and curation strategies with specific downstream tasks and the inherent characteristics of multimodal inputs, dataset construction can implicitly guide the model in learning more relevant and robust representations. NuScenes-SpatialQA, the first large-scale benchmark designed to evaluate the spatial understanding and reasoning capabilities of VLMs in autonomous driving scenarios. The authors found that while VLMs perform reasonably well in qualitative spatial tasks, they struggle significantly with quantitative reasoning \cite{tian2025spatialQA}. NuPlanQA dataset offers a large-scale collection of 1 million real-world VQA pairs across nine subtasks, providing a comprehensive benchmark for evaluating MLLMs in complex driving scenarios with diverse and free-form questions. This combination addresses critical limitations in existing approaches by improving both the input representation and the evaluation methodology for autonomous driving MLMs \cite{park2025nuplanqa}. However, the principles of dataset construction remain largely empirical, often guided by experience rather than systematic methodology. The process is time-consuming and labor-intensive and requires extensive data collection, synchronization, and annotation.

With these insights, we propose to take advantage of the closely coupled nature of sensor modalities in autonomous driving to design more effective VLAs. Exploiting these inherent cross-modal correspondences, our approach enhances model learning with minimal hand-crafted design.

\section{Methodology}
\begin{figure*}
    \centering
    \begin{subfigure}{.47\linewidth}
        \centering
        \includegraphics[width=\linewidth]{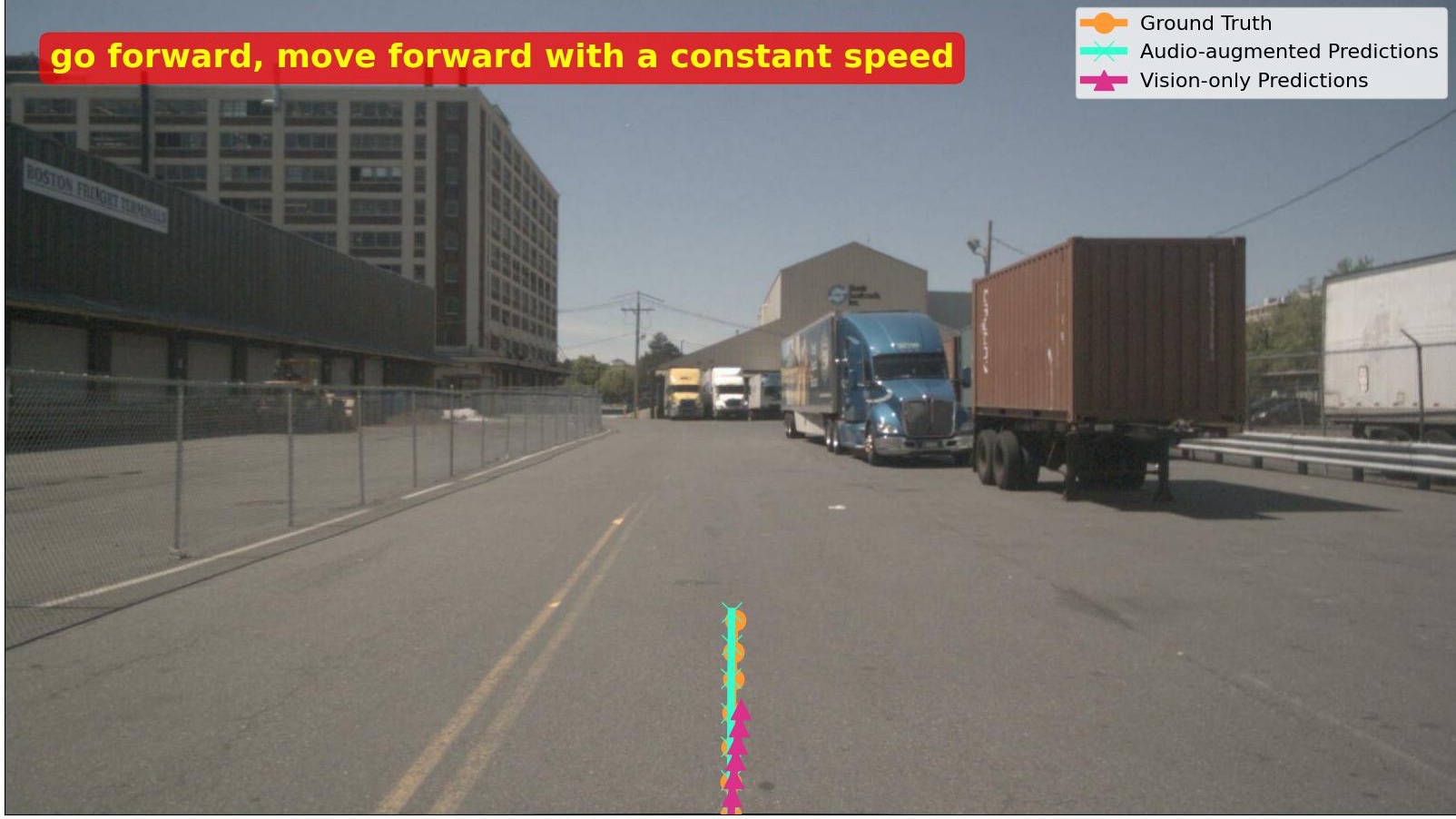}
    \end{subfigure}
    \begin{subfigure}{.47\linewidth}
        \centering
        \includegraphics[width=\linewidth]{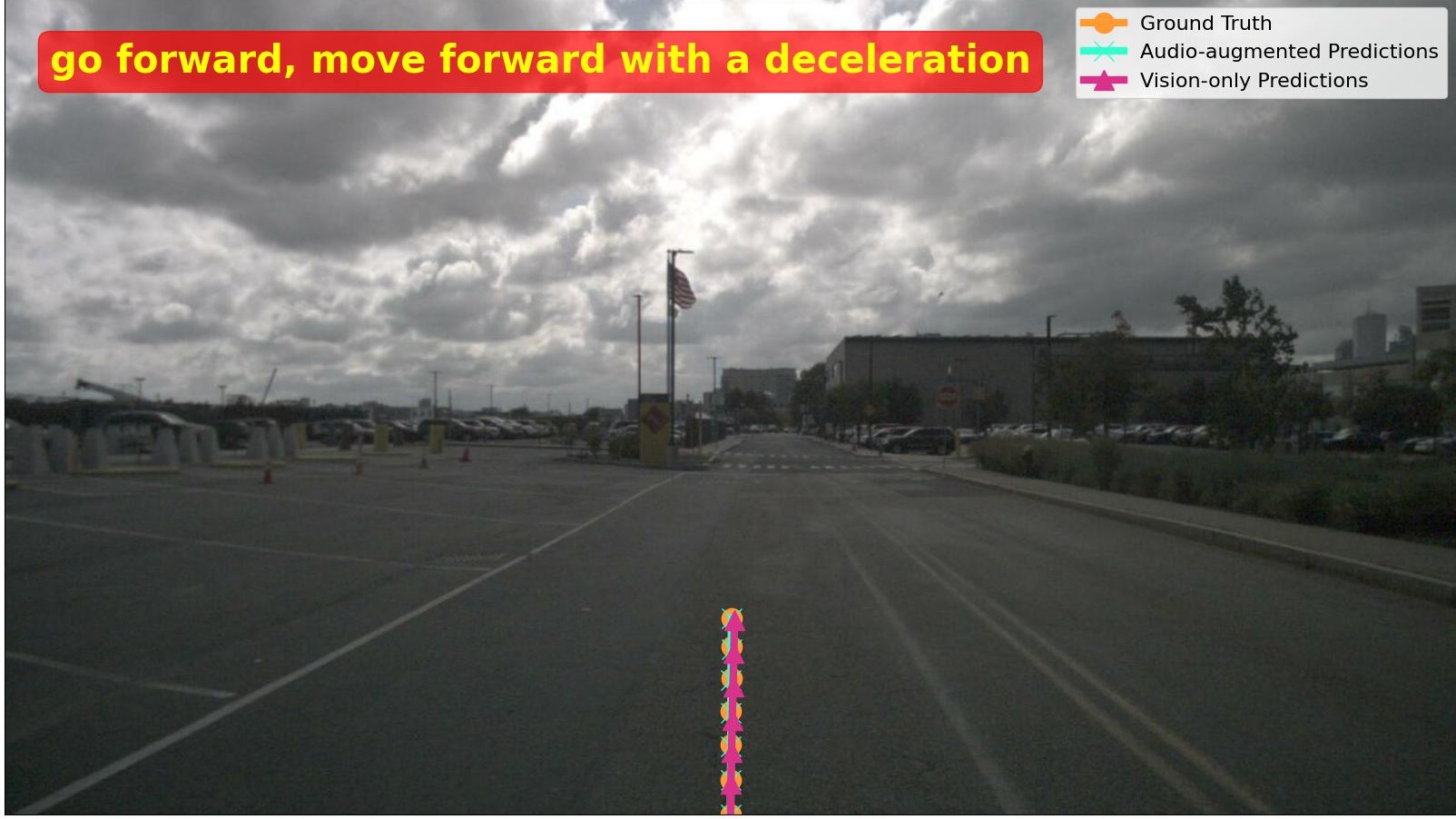}
    \end{subfigure}
    \caption[width=.7\linewidth]{In driving scenarios that afford limited maneuvering options, vision-based perception can generate contextually appropriate actions.}
    \label{fig:vis_well}

    \vspace{2mm}
    
    \begin{subfigure}{.47\linewidth}
        \centering
        \includegraphics[width=\linewidth]{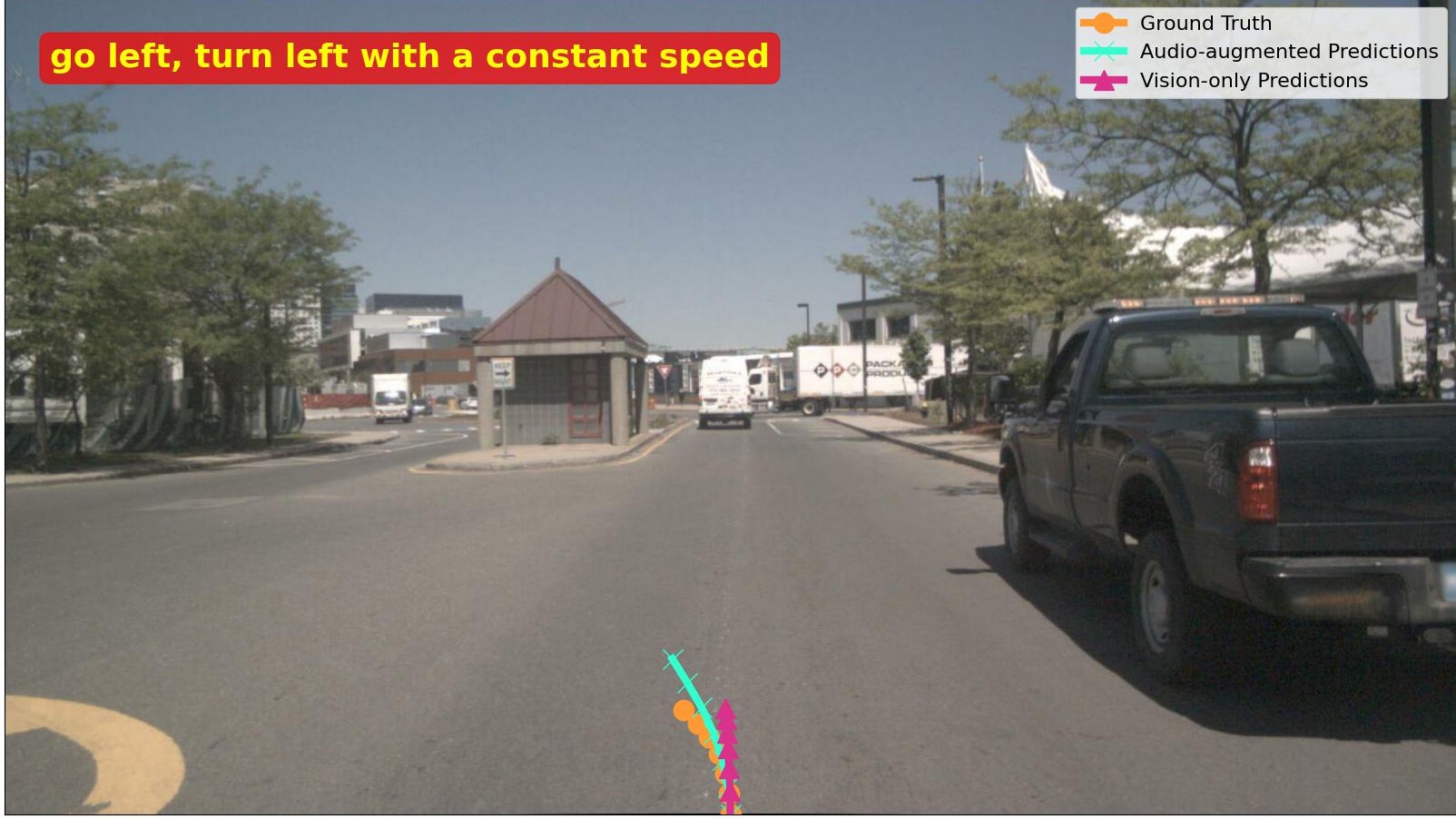}
    \end{subfigure}
    \begin{subfigure}{.47\linewidth}
        \centering
        \includegraphics[width=\linewidth]{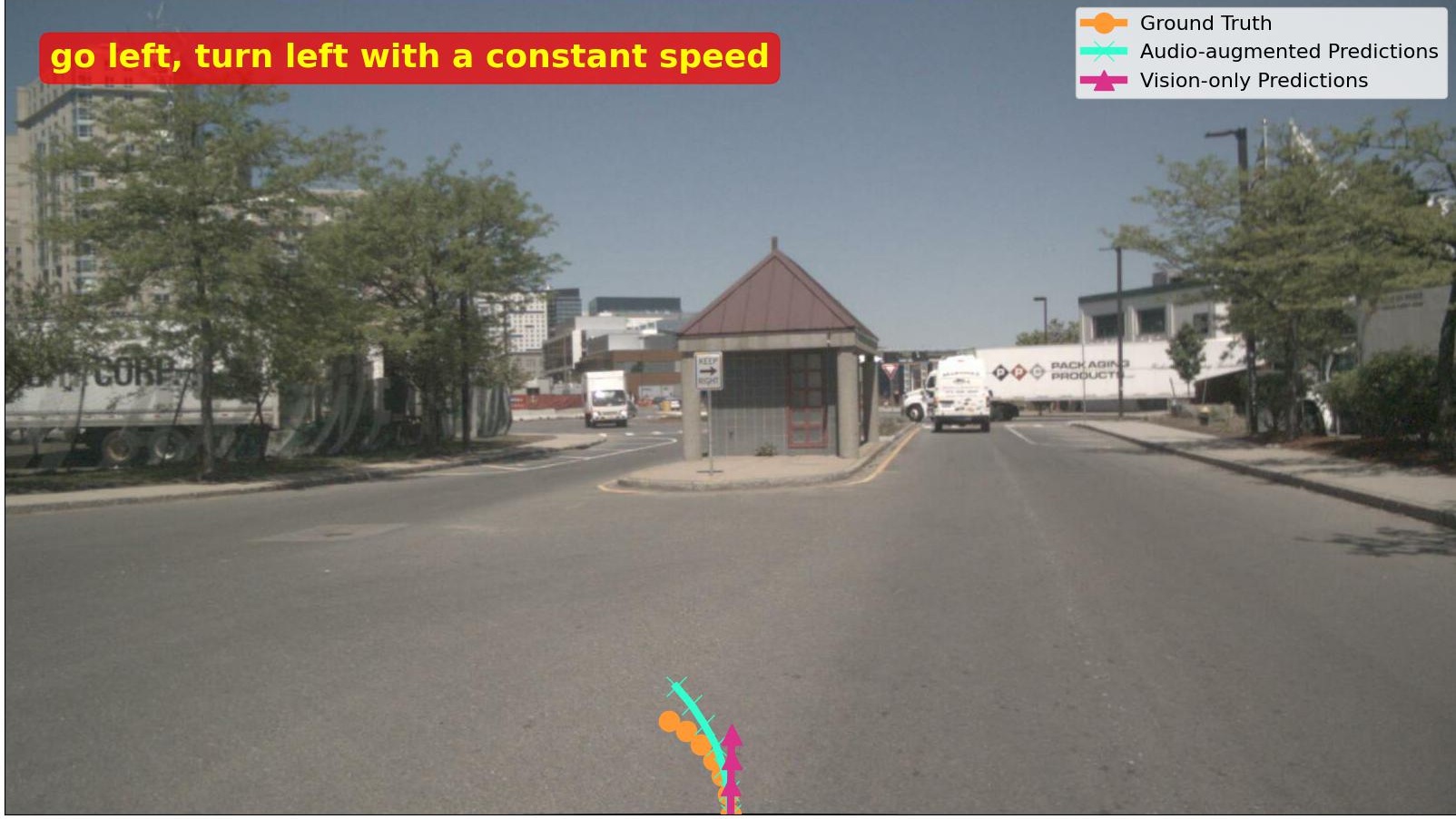}
    \end{subfigure}
    \caption[width=.7\linewidth]{In driving scenarios that present multiple viable options, the user’s intention may diverge from the trajectory implied by current visual observations. In such cases, audio-guided trajectories adhere to the user command, yielding plans that deviate from those produced by a visual observation-only VLA.}
    \label{fig:vis_compare}
\vspace{-4mm}
\end{figure*}

\subsection{Multimodal CoT}
Based on nuScenes, we process the history ego trajectory and future ego trajectory with k-means clustering for each frame. Then we obtain the classification of the intentions of the ego vehicle. Meanwhile, analyzing the ego states including speed and acceleration, we translate the ego states into a structured natural language $\{\mathbf{Goal}, \mathbf{Current\_action}\}$. With the Text-To-Speech model, we convert structured texts to audio commands \cite{ren2020fastspeech}. 

To synthesize emotion-augmented audio commands, we alter the speech tempo and pitch of the base audio to create additional emotional audio commands with the same audio content \cite{yang2025emovoice}. For each audio command, we compute the arousal that quantifies the perceived energy of the audio and label the audios with urgent and hesitant emotions accordingly. The arousal ranges from $0$ to $1$, where a high arousal of an audio denotes an urgent voice and a low arousal represents a hesitant voice \cite{bone2014robust}. We compute arousal as a weighted sum of sigmoid-transformed, normalized audio features. The sigmoid transformation is defined as
\begin{equation}
    \sigma(x; k, x_0) = \frac{1}{1 + \exp(-k(x - x_0))}.
\end{equation}
Then the arousal is computed as
\begin{equation}
    \begin{split}
    A = 0.4 \cdot \sigma(R_n; 8, 0.4) +
    0.4 \cdot \sigma(F_n; 10, 0.5) + \\
    0.15 \cdot \sigma(T_n; 7, 0.5) +
    0.05 \cdot \sigma(C_n; 6, 0.4)
    \end{split}
\end{equation}
where $R_n$, $F_n$, $T_n$, and $C_n$ are the normalized mean RMS energy, fundamental frequency (F0), tempo (BPM) and spectral centroid of an audio signal \cite{kossaifi2019sewa}. Then the arousal $A$ is clipped to
\begin{equation}
    A = \min(1, \max(0, A)).
\end{equation}

Based on the ratios between the synthetic and original arousal values, we modulate the corresponding trajectories with a faster or slower speed profile without changing the original driving directions. Given an input trajectory $\mathbf{P} = \{\mathbf{p}_i \in \mathbb{R}^2\}_{i=0}^{N-1}$ sampled uniformly over time $t \in [0, T]$ and a target emotion $e \in \{\text{urgent}, \text{hesitant}\}$, we define the modulation procedure to synthesize a new trajectory $\tilde{\mathbf{P}} = (\tilde{p}_x, \tilde{p}_y)$ according to an emotion-conditioned speed profile as follows.

First, we compute the cumulative Euclidean path length $\mathbf{L}$,
\begin{equation}
    \Delta \mathbf{p}_i = \mathbf{p}_{i+1} - \mathbf{p}_i, \quad i = 0, \dots, N-2
\end{equation}
\begin{equation}
    d_i = \| \Delta \mathbf{p}_i \|_2
\end{equation}
\begin{equation}
    L_0 = 0, \quad L_i = \sum_{k=0}^{i-1} d_k, \quad i = 1, \dots, N-1.
\end{equation}
\begin{equation}
    \mathbf{L} = [L_0, L_1, \dots, L_{N-1}]^\top
\end{equation}
where $N$ is the number of waypoints in the trajectory. Then the total path length is $L_{\text{total}} = L_{N-1}$.

We compute the base average speed from the original trajectory. This serves as a reference speed for emotion-based modulation.
\begin{equation}
    v_{\text{avg}} = \frac{L_{\text{total}}}{T}
\end{equation}
where $T$ is the total duration of a trajectory in seconds.

We then define $v_e(t) \in \mathbb{R}^+$ as the emotion-conditioned speed function. For urgent emotion,
\begin{equation}
    v_{\text{urgent}}(t) = v_{\text{avg}} \cdot 1.6 \cdot \left(1 - e^{-2t}\right) + \epsilon(t)
\end{equation}
where $\epsilon(t) \sim \mathcal{N}(0, (0.03 v_{\text{avg}})^2)$, and we clip $v_{\text{urgent}}(t)$ to $v_{\text{urgent}}(t) \in [0.8 v_{\text{avg}},\; 1.5 v_{\text{avg}}]$.

For hesitant emotion, 
\begin{equation}
    v_{\text{hesitant}}(t) = v_{\text{avg}} \cdot 1.2 \cdot \left(1 - e^{-t}\right) + \epsilon(t)
\end{equation}
where $\epsilon(t) \sim \mathcal{N}(0, (0.05v_{\text{avg}})^2)$, and we perform hesitation modification at midpoint:
\begin{equation}
    v_{\text{hesitant}}(t_i) =
    \begin{cases}
        0.6 \cdot v_{\text{hesitant}}(t_i), & \text{if } |i - \lfloor N/2 \rfloor| \leq 1 \\
        v_{\text{hesitant}}(t_i), & \text{otherwise}.
    \end{cases}
\end{equation}

Then we clip $v_{\text{hesitant}}(t)$ to $v_{\text{hesitant}}(t) \in [0.1 v_{\text{avg}},\; 1.5 v_{\text{avg}}]$

Second, we perform reparameterization by the speed profiles, where we compute how far along the path the ego vehicle should be at each time step.
\begin{equation}
    \Delta t_i = t_{i+1} - t_i, \quad i = 0, \dots, N-2
\end{equation}
Having the interval $\Delta t_i$, the average speed per interval is defined as
\begin{equation}
    \bar{v}_i = \frac{v_e(t_i) + v_e(t_{i+1})}{2}.
\end{equation}

Then we define the segment distance as
\begin{equation}
    \delta s_i = \bar{v}_i \cdot \Delta t_i
\end{equation}
   
So, the cumulative distance $\mathbf{S}$ is calculated as
\begin{equation}
    S_0 = 0, \quad S_{i+1} = S_i + \delta s_i, \quad i = 0, \dots, N-2
\end{equation}
\begin{equation}
    \mathbf{S} = [S_0, S_1, \dots, S_{N-1}]^\top
\end{equation}
We then normalize $\mathbf{S}$ to match the total path length:
\begin{equation}
    \tilde{\mathbf{S}} = L_{\text{total}} \cdot \frac{\mathbf{S}}{\max(\mathbf{S})}
\end{equation}
    
With clamping to valid range, 
\begin{equation}
     \hat{\mathbf{S}} = \min(\max(\tilde{\mathbf{S}}, 0), L_{\text{total}}).
\end{equation}
Finally, we interpolate new positions along the path. For each dimension $x$ and $y$,
\begin{equation}
\begin{aligned}
    \tilde{p}_x(S) =& \sum_{i=0}^{N-2} \mathbf{I}_{[L_i, L_{i+1})}(S) \\ &\left(p_{x,i} + \frac{S - L_i}{L_{i+1} - L_i}(p_{x,i+1} - p_{x,i})\right)
\end{aligned}
\end{equation}
where $\mathbf{I}_{[L_i, L_{i+1})}(S)$ is an indicator function that equals 1 if $S$ is within the interval $[L_i,L_{i+1})$ and 0 otherwise. Similarly for $\tilde{p}_y(S)$. Fig. \ref{fig:emo_traj} shows the modulation of trajectories according to emotion labels. Given the front-view images and user audios, we construct the CoT inferring the arousal and emotion of audio and the corresponding trajectory with modulated speed profile. In Fig. \ref{fig:dataset}, we show an example of our multimodal CoT dataset.

\subsection{Supervised Fine-tuning on Qwen2.5-Omni}

In multimodal tasks, especially those that involve both audio and image inputs, a key challenge lies in effectively performing cross-modal temporal alignment. Given that audio signals typically have higher temporal resolution compared to images or video frames, which are sampled at much lower frequencies, there exists a significant disparity in time scales between these two modalities. Qwen2.5-Omni fuses multiple modalities with Time-aligned Multimodal RoPE (TMRoPE), a novel rotary embedding that interleaves audio and image frames to guarantee temporal alignment, while its Thinker-Talker architecture dedicates a Transformer decoder (Thinker) to high-level understanding and a dual-track autoregressive decoder (Talker) that receives Thinker’s latent state \cite{xu2025qwen2_5}. 

In Fig. \ref{fig:overview}, we show our proposed framework, where Thinker decodes the audio command, camera image, and system prompt embeddings, and a text head outputs the audio analysis, emotion detection, and trajectory waypoints based on CoT reasoning. Taking into account the characteristics of long-sequence audio input, Qwen2.5-Omni implements efficient streaming processing capabilities. Specifically, both audio and visual encoders support block-wise processing and utilize sliding-window attention mechanisms to reduce latency. With our multimodal CoT dataset, our supervised fine-tuning not only overcomes the inherent temporal asymmetry between audio and image inputs, but also provides consistent action output based on audio instructions and emotion detections.

\section{Experiments}
\balance
\begin{figure}
    \centering
    \includegraphics[width=0.83\linewidth]{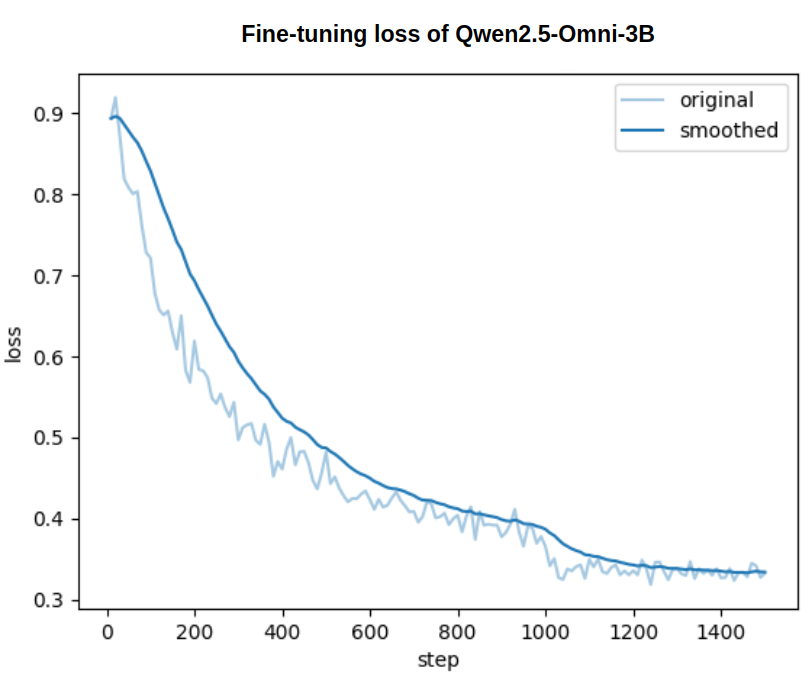}
    \caption{Fine-tuning loss decreases smoothly during the experiment.}
    \label{fig:loss}
    \vspace{-5mm}
\end{figure}

\subsection{Experimental Setup}

The nuScenes dataset \cite{caesar2020nuscenes} is chosen for training and evaluation of our framework, which is a large-scale dataset for autonomous driving, consisting of 1000 scenes with measurements from 1 spinning LiDAR, 6 cameras and 5 long-range radars, etc. LiDAR has 20 Hz capture frequency and 32 beams, while cameras have 12 Hz capture frequency and $1600 \times 900$ resolution. Radars are with 13 Hz capture frequency. The well-synchronized keyframes from the cameras, LiDAR, and radars are at 2 Hz.

\subsection{Supervised Fine-tuning Recipe}

Fine-tuning was performed on four Nvidia RTX 3090 24G GPUs with learning rate $1e-05$, training batch size $1$, and evaluation batch size $8$. We used AdamW optimizer with $\beta=(0.9,0.999)$ and $\epsilon=1e-08$ and cosine scheduler with scheduler warm-up ratio $0.1$. Fine-tuning was performed in 3 epochs, where we show the loss-step plot in Fig. \ref{fig:loss}. As we can see, during fine-tuning, the VLM exhibits a stable and consistently decreasing cross-entropy loss curve, with no significant oscillations or plateaus, indicating effective optimization and smooth convergence on the nuScenes training set.

\begin{figure*}[]
    \centering
    \begin{subfigure}{0.496\linewidth}
        \centering
        \includegraphics[width=0.99\linewidth, keepaspectratio]{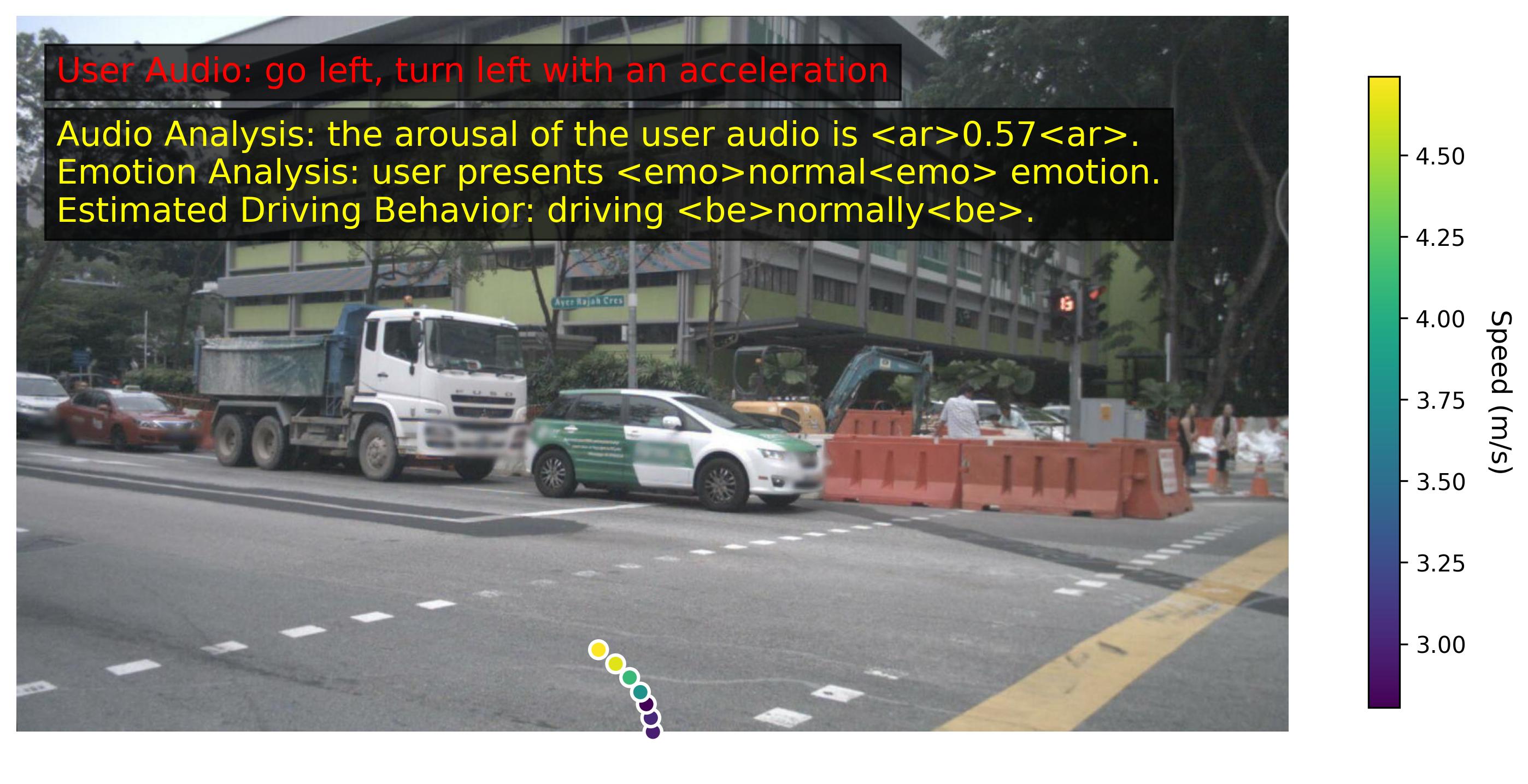}
    \end{subfigure}
    \begin{subfigure}{0.496\linewidth}
        \centering
        \includegraphics[width=0.99\linewidth, keepaspectratio]{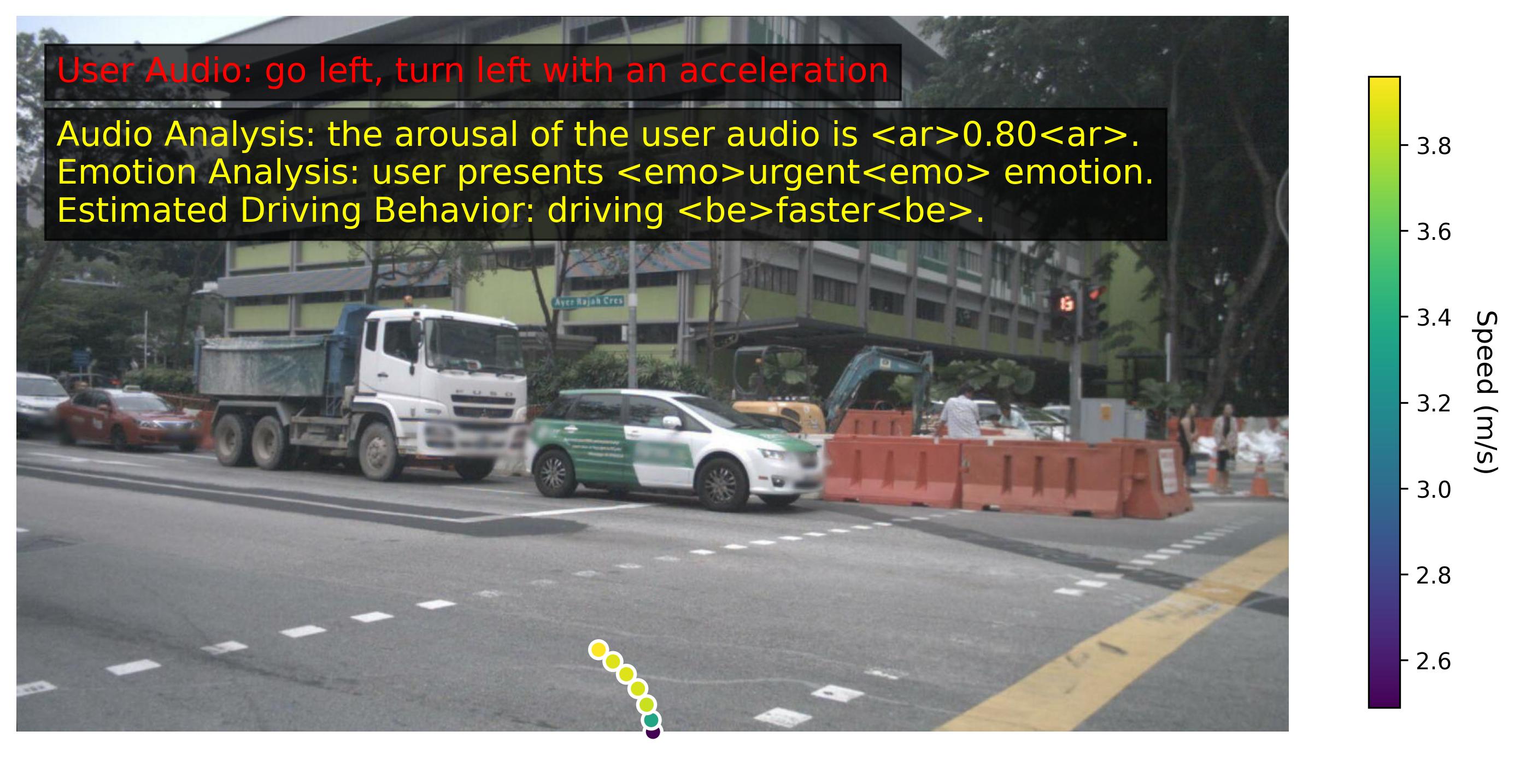}
    \end{subfigure}
    \hfill
    \begin{subfigure}{0.496\linewidth}
        \centering
        \includegraphics[width=0.99\linewidth, keepaspectratio]{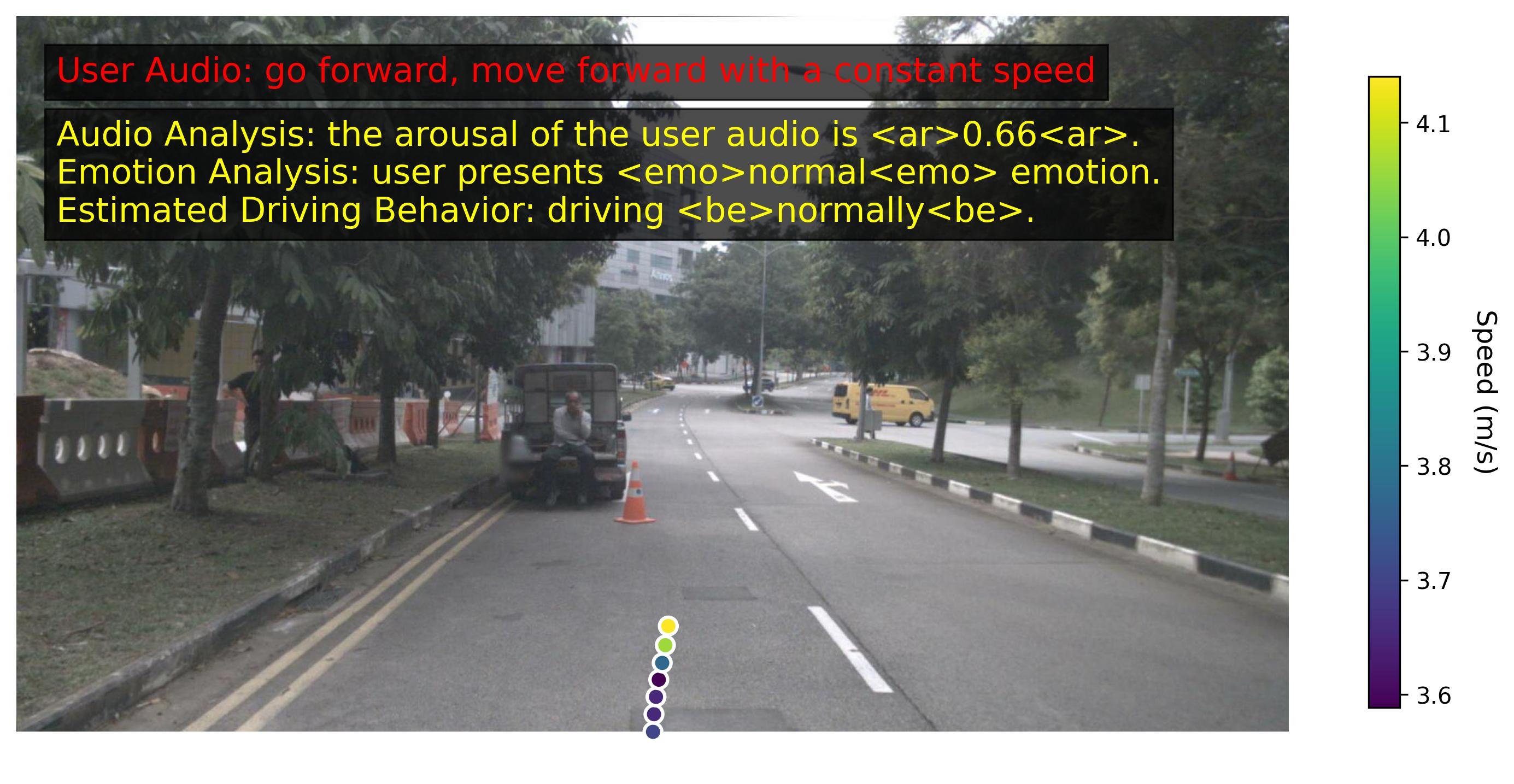}
    \end{subfigure}
    \begin{subfigure}{0.496\linewidth}
        \centering
        \includegraphics[width=0.98\linewidth, keepaspectratio]{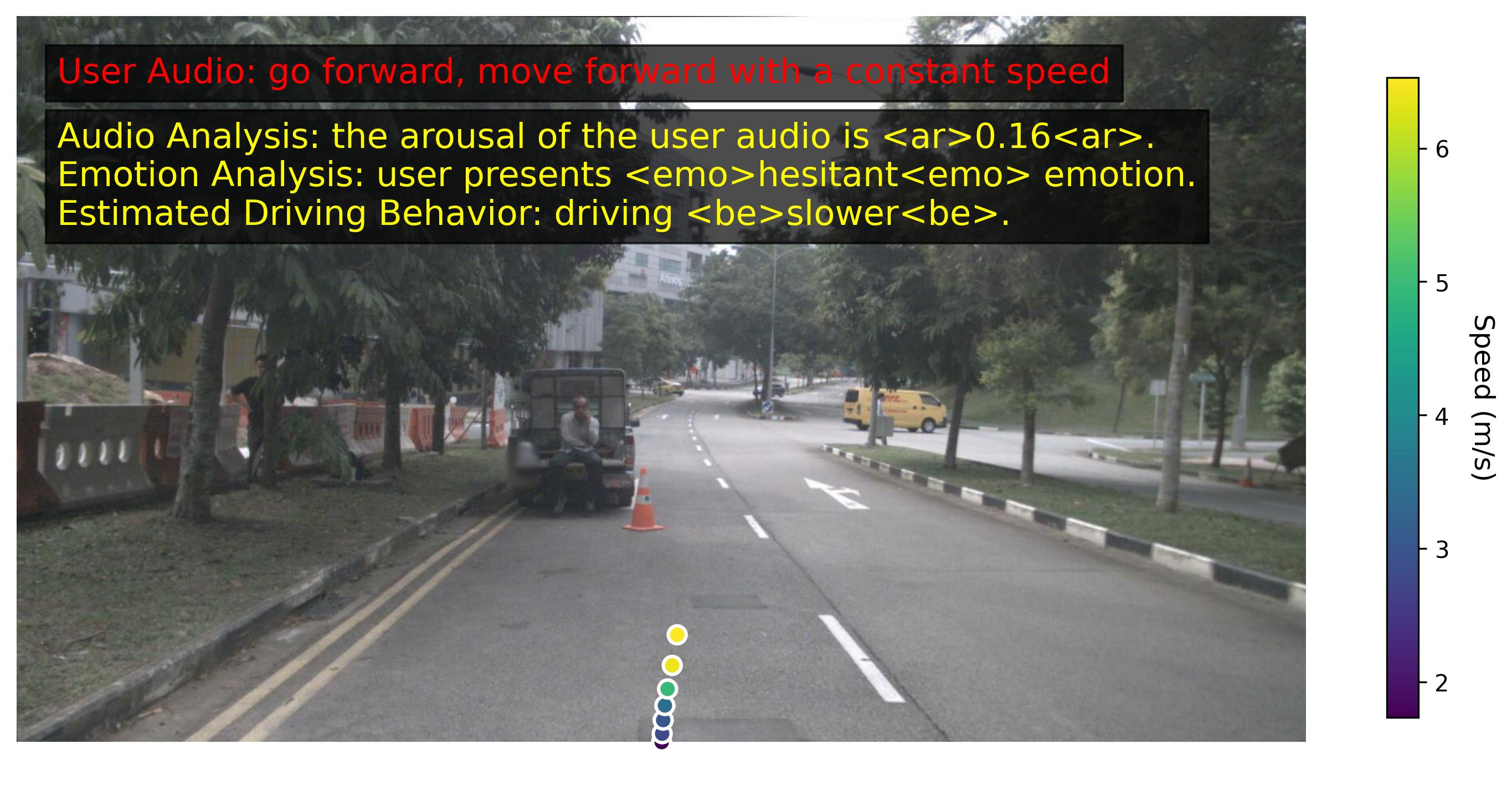}
    \end{subfigure}
    \caption{The results of EchoVLA's emotion analysis and driving behavior guidance, where we show the speed at each waypoint via color bar. Based on the CoT thinking, the driving behavior is adjusted with corresponding speed profiles. The left column shows the trajectory prediction based on normal user emotion and the right column shows the trajectory prediction based on urgent or hesitant emotion detection in the same scenes.}
    \label{fig:emo_record}
    \vspace{-4mm}
\end{figure*}

\subsection{Evaluation Methods}

We demonstrate our fine-tuned VLA's performance by the L2 error in meters and the collision rate in percentage. The average L2 error is computed as the distance between each waypoint in the planned trajectory and its corresponding waypoint in the ground truth trajectory, serving as a measure of how closely the planned trajectory matches a human-driven one. The collision rate is evaluated by placing an ego-vehicle bounding box at each waypoint along the planned trajectory and checking for overlaps with the ground truth bounding boxes of other objects.

To validate our proposed method, we compared our EchoVLA with visual perception-only VLA on the nuScenes open-loop L2 error and collision rate. The baseline visual perception-only VLA is based on a larger VLM Qwen2-VL-7B compared to our 3B scale model. In Qwen2-VL, Multimodal Rotary Position Embedding (M-RoPE) is applied to process multimodal input by decomposing rotary embedding into temporal, height, and width components, which equips Qwen2-VL with robust multimodal data handling capabilities \cite{wang2024qwen2vl}. The baseline visual perception-only VLA is also fine-tuned on the nuScenes dataset.

In Table \ref{tab:L2}, by adding audio supervision, our method outperforms the visual perception-only baseline with a $59.4\%$ lower L2 error and $74.4\%$ lower collision rate in the nuScenes validation set.

By showing the audio with normal, urgent and hesitant moods, in Fig. \ref{fig:emo_record}, we report that the changes in tones, with the same audio content, alter speed, braking, and steering, proving that the mood signal guides how the car drives according to the user's intention.

\subsection{Qualitative Results}

In Fig. \ref{fig:vis_well} and Fig. \ref{fig:vis_compare}, we show the visualization of the ground truth trajectory, visual perception-only, and our EchoVLA's trajectory prediction from the samples of the nuScenes dataset, where the input audio commands are placed in the left upper corner and the visual perception-only VLA is based on Qwen2-VL-7B. In Fig. \ref{fig:vis_well}, both visual perception-only and our EchoVLA predict reasonable trajectories, as the driving intention is encoded identically by them. However, in Fig. \ref{fig:vis_compare}, when the audio command poses a different driving intention compared to visual understanding, a visual perception-only VLA cannot follow the actual driving command. Our EchoVLA is capable of guiding the driving with user intention-aware trajectory output based on the audio input of the user.  

\subsection{Ablation Study}

\begin{table*}[t]
\caption{The open-loop planning comparison on nuScenes evaluation set.}
\centering
\resizebox{0.9\linewidth}{!}{
    \begin{tabular}{l|l|c|c|c|c|c|c|c|ccc}
        \toprule
        \multirow{2}{*}{Perception Modality} & \multirow{2}{*}{Model} & \multicolumn{4}{c|}{L2 (m) $\downarrow$} & \multicolumn{4}{c}{Collision Rate (\%) $\downarrow$} \\
         & & 1s & 2s & 3s & Avg. & 1s & 2s & 3s & Avg. \\
        \midrule
        Vision & Qwen2-VL-7B & 0.55 & 1.20 & 2.54 & 1.43 & 0.06 & 0.17 & 1.07 & 0.43 \\  
        \midrule
        Vision & VLP-UniAD & 0.36 & 0.68 & 1.19 & 0.74 & 0.03 & \textbf{0.12} & 0.32 & 0.16 \\   
        \midrule
        Vision & RDA-Driver & 0.23 & 0.73 & 1.54 & 0.80 & \textbf{0.00} & 0.13 & 0.83 & 0.32 \\ 
        \midrule
        
        Vision + Audio & \textbf{Ours} & 0.46 & \textbf{0.52} & \textbf{0.74} & \textbf{0.58} & \textbf{0.00} & \textbf{0.12} & \textbf{0.22} & \textbf{0.11} \\
        \midrule
    \end{tabular}}
\label{tab:L2}
\vspace{-5mm}
\end{table*}

In this section, we show the ablation study of different approaches to the embedding of audio signals for a VLM. In Table \ref{tab:ablation}, we use HuBERT, VQ-VAE, and STFT to encode audio to obtain embeddings, where audio embeddings are treated as extra tokens for VLM's encoders. In HuBERT, the audio is encoded as a feature representation of a temporal sequence \cite{hsu2021hubert}. Through VQ-VAE, the audio is encoded as the sequence of discrete codes \cite{van2017vqvae}, while the audio is treated as a complex-valued matrix in STFT, where one axis represents time and the other axis represents frequency bins \cite{feng2025stftcodec}. The fine-tunings in the ablation study are performed in the nuScenes mini set, where we demonstrate that our approach with inherent MLM encoding and CoT reasoning is the most effective.

\begin{table}[h]
\caption{Ablation study of different approaches of embedding the audio signals on nuScenes mini set.}
\centering
\resizebox{0.9\linewidth}{!}{
\begin{tabular}{|c|c|c|}
\hline
Methods & Avg. L2 Error & Avg. Collision Rate \\
\hline
huBert \cite{hsu2021hubert} & 0.45 & 0.33 \\
\hline
VQ-VAE  \cite{van2017vqvae} & 0.48 & 0.32 \\
\hline
STFT  \cite{feng2025stftcodec} & 0.60 & 0.46 \\
\hline
\textbf{Ours}  & \textbf{0.38} & \textbf{0.24} \\
\hline
\end{tabular}}
\label{tab:ablation}
\vspace{-3mm}
\end{table}

\section{Conclusion}
We introduce EchoVLA, which augments VLA with an online audio channel, allowing users to inject time-varying intent without hand-engineered reward reshaping or additional advanced sensors. By converting ego-motion descriptions into synchronized speech commands and fine-tuning a Qwen2.5-Omni backbone, EchoVLA learns to align acoustic intent with visual context and regresses trajectories in a CoT reasoning way. The open-loop evaluation in nuScenes confirms that audio enhancement is complementary to, rather than competing with, visual perception. 

However, current audio command generation relies on synthesized speech derived from recorded ego trajectories, which does not fully capture the variability of natural human speech. In practice, user utterances are often noisier, span multiple languages, and can be ambiguous in meaning. Additionally, the model processes the entire audio clip before generating a response, increasing latency for longer commands—streaming token generation has not yet been implemented. Our system assumes clean audio input and has not been evaluated under realistic acoustic conditions such as heavy traffic noise, wind, or background music playing inside the cabin.

Moving forward, we plan to address these limitations by curating a large-scale corpus of real in-car voice commands that reflect diverse languages, accents, and paraphrased expressions. We also intend to enhance robustness through an adversarial noise-training phase and by incorporating an on-board denoising front-end to handle challenging acoustic environments.

Our approach reduces the average L2 error by $59.4\%$ and the collision rate by $74.4\%$ compared to the baseline VLA with vision-only perception. We hope that this work establishes a milestone in improving the safety of VLAs in autonomous driving by efficiently integrating multiple modalities.

\bibliographystyle{IEEEtran}
\balance
\bibliography{references}
\end{document}